\documentclass[pre, twocolumn, superscriptaddress, amsmath, amssymb, floatfix, preprintnumbers]{revtex4-2}
\usepackage[utf8]{inputenc} 
\usepackage[T1]{fontenc}
\usepackage[latin9]{luainputenc}
\usepackage{babel}
\usepackage{amsbsy}
\usepackage{amstext}
\usepackage{dcolumn}
\usepackage{color}
\usepackage{lipsum}

\usepackage{dsfont}
\usepackage{dirtytalk}
\usepackage{graphicx}
\usepackage{multirow}
\usepackage{setspace}
\usepackage{float}
\usepackage[pdfusetitle, bookmarks=true,bookmarksnumbered=false,bookmarksopen=false, breaklinks=false,pdfborder={0 0 1},backref=false,colorlinks=false]{hyperref}
\hypersetup{colorlinks = true, linkcolor = blue, anchorcolor = blue, citecolor = blue, filecolor = blue, urlcolor = blue}

\newcommand{\orcidlink}[1]{\href{https://orcid.org/#1}{\includegraphics[width=10pt]{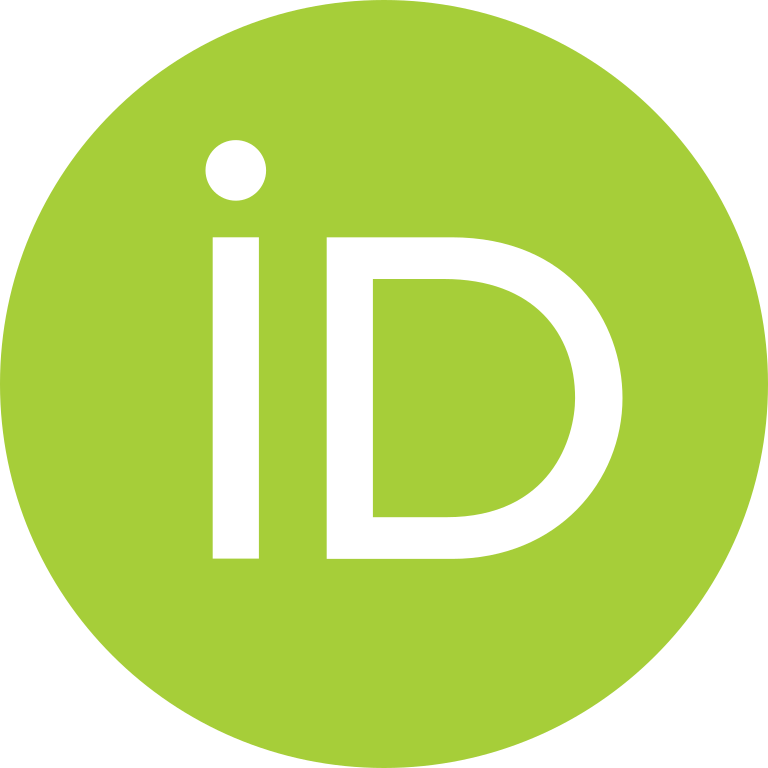}}}
 
\begin{document}



\title{Emergent Competition Between Dynamical Channels in Nonequilibrium Systems}

\author{R. A. Dumer \orcidlink{0000-0002-9032-111X}}
\email{rafaeldumer@fisica.ufmt.br}
\affiliation{Programa de Pós-Graduação em Física, Instituto de Física, Universidade Federal de Mato Grosso, Cuiabá, Brazil}
\affiliation{Department of Physics \& I3N, University of Aveiro, 3810-193 Aveiro, Portugal}

\author{M. Godoy \orcidlink{0000-0001-9122-6061}}
\email{mgodoy@fisica.ufmt.br}
\affiliation{Programa de Pós-Graduação em Física, Instituto de Física, Universidade Federal de Mato Grosso, Cuiabá, Brazil}

\author{J. F. F. Mendes \orcidlink{0000-0002-4707-5945}}
\email{jfmendes@ua.pt}
\affiliation{Department of Physics \& I3N, University of Aveiro, 3810-193 Aveiro, Portugal}

\begin{abstract}
We introduce a rejection-free continuous-time kinetic Monte Carlo framework to study stochastic systems governed by multiple concurrent dynamical mechanisms. In this approach, the relative activity of each dynamical channel emerges self-consistently from the instantaneous configuration through its transition rates. As an illustration, we investigate a driven antiferromagnetic Ising model on a square lattice combining conservative Katz–Lebowitz–Spohn exchanges and nonconserving Glauber single-spin flips. We show that the coexistence of these dynamics qualitatively reshapes the nonequilibrium phase diagram in the temperature–field plane, stabilizing antiferromagnetic order in regions where the driving field would otherwise destroy it. Near the zero-temperature limit, the phase boundary follows a power-law scaling $T\sim|E-E_c|$ with an exponent close to unity. At intermediate temperatures, the transition belongs to the two-dimensional Ising universality class, while at low temperatures it remains continuous, with the order-parameter exponent approaching zero. Our results demonstrate that allowing competing dynamical channels to coevolve with the system can fundamentally alter its critical properties, revealing collective behavior hidden in single-dynamics descriptions.
\end{abstract}

\maketitle

Many physical systems are governed microscopically by more than one concurrent mechanism. Representative examples include charge transport with recombination in semiconductors~\cite{ShockleyRead1952,Hall1952}, ion intercalation with finite charge-transfer kinetics in batteries~\cite{DoyleFullerNewman1993,Bazant2013}, and magnon spin transport coexisting with damping and spin--lattice relaxation in magnetic materials~\cite{Tserkovnyak2002,Chumak2015,Cornelissen2015}. In broad terms, these settings often feature a conservative transport channel that redistributes a locally conserved quantity, together with a nonconservative relaxational channel that exchanges energy (or an order parameter) with an effective bath~\cite{HohenbergHalperin1977,Glauber1963,Kawasaki1966a,Kawasaki1966b,Parmeggiani2003,Evans2005}.

A recurring challenge in this broad class of nonequilibrium systems is to determine how frequently each mechanism is actually executed in the steady state, how each channel shapes the transient approach to stationarity, and how their relative importance depends on the instantaneous configuration. A common simplification is to prescribe fixed attempt frequencies (or a fixed mixing parameter) for the different updates~\cite{PRA1989,arek2025,PRE2000,PRE1999,PRE2025}, however, this can obscure the genuinely concurrent and configuration-dependent character expected when distinct mechanisms operate simultaneously.

In this Letter, we introduce a natural framework for treating multi-channel stochastic dynamics with well-defined transition rates, based on rejection-free, continuous-time kinetic Monte Carlo (CTKMC). Here, channels refer to distinct physical mechanisms (dynamical rules) acting on the same configuration space. Within this framework, we show, using a concrete example, that the nonequilibrium stationary state, critical behavior, and even the universality class may change substantially once an additional mechanism is switched on and allowed to coevolve with the system.

In our formulation of rejection-free CTKMC~\cite{FrontChem2019,Gillespie1977,BKL1975}, each channel $a$ is specified by well-defined transition rates $w^{a}_{\boldsymbol{\sigma}\to \boldsymbol{\sigma}'}$. The associated channel-resolved escape rate from the current configuration $\boldsymbol{\sigma}$ is
\begin{equation}
W_a(\boldsymbol{\sigma})=\sum_{\boldsymbol{\sigma}^\prime\neq \boldsymbol{\sigma}} w^{a}_{\boldsymbol{\sigma}\to \boldsymbol{\sigma}'} ,
\qquad
W(\boldsymbol{\sigma})=\sum_a W_a(\boldsymbol{\sigma}),
\label{eq:1}
\end{equation}

\begin{figure}
\begin{centering}
\begin{centering}
\includegraphics[scale=0.17]{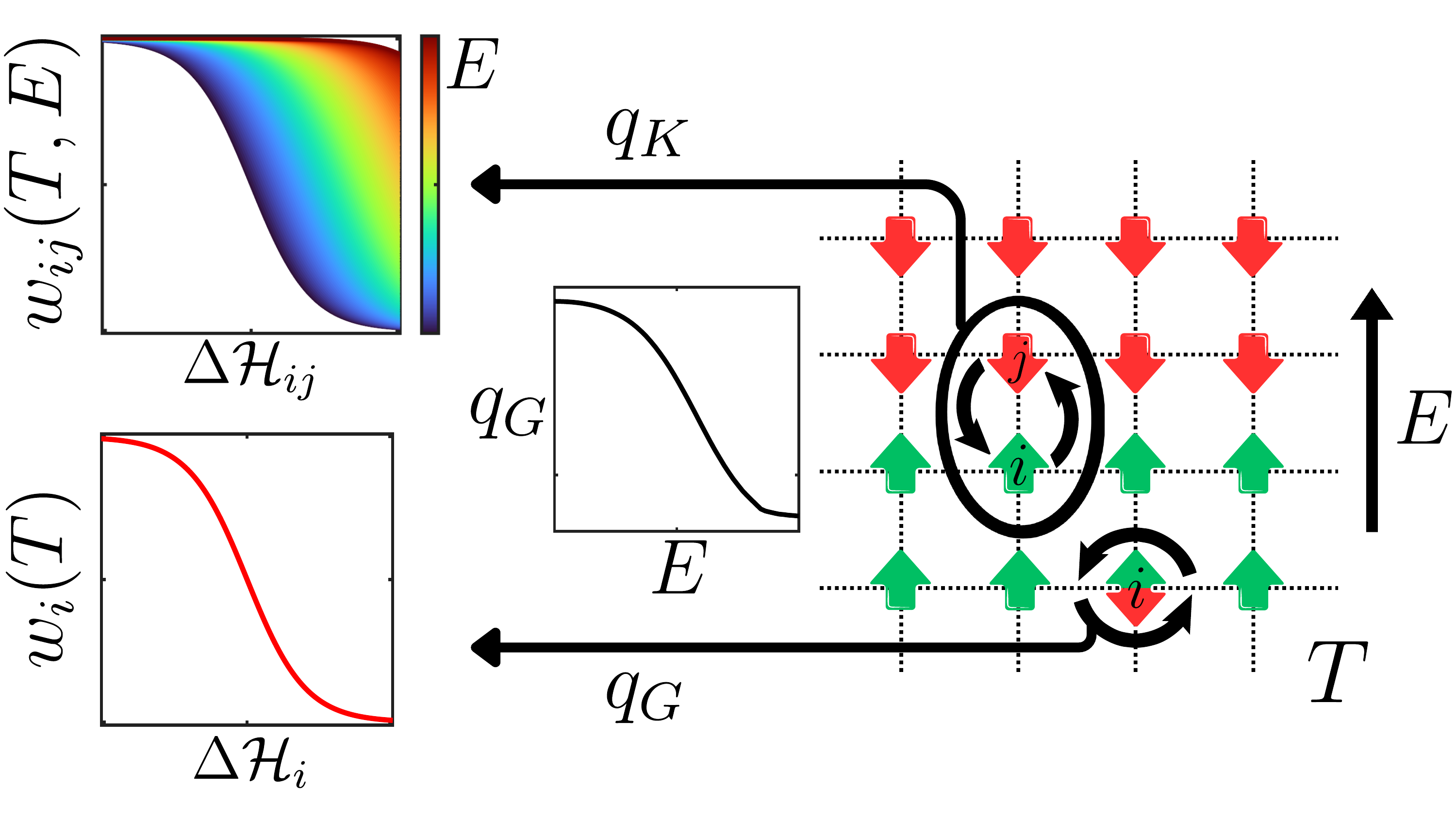}
\end{centering}
\caption{\textcolor{blue}{(Color online)} Schematic illustration of a system under the influence of two competing dynamical mechanisms. The conservative channel (two-spin exchange) has transition rate $w_{ij}$, which depends on the external parameters drive $E$ and temperature $T$, as well as on the internal energy variation $\Delta\mathcal{H}_{ij}$, and acts with probability $q_K = 1 - q_G$ determined by the instantaneous configuration. In contrast, the nonconserving channel (single-spin flip) has transition rate $w_i$, which depends on $T$ and on the internal energy variation $\Delta\mathcal{H}_{i}$, and acts with probability $q_G$, also determined by the current state of the system.}
\label{fig:1-1}
\end{centering}
\end{figure}
where $W(\boldsymbol{\sigma})$ is the total escape rate. An event is generated by drawing an exponentially distributed waiting time $\Delta t$ with mean $1/W(\boldsymbol{\sigma})$ (equivalently, $\Delta t=-\ln u/W(\boldsymbol{\sigma})$ with $u\in(0,1)$ uniform), selecting the channel with probability $q_a=W_a(\boldsymbol{\sigma})/W(\boldsymbol{\sigma})$, and then selecting the specific transition $\boldsymbol{\sigma}\to\boldsymbol{\sigma}'$ within the chosen channel with probability $w^{a}_{\boldsymbol{\sigma}\to\boldsymbol{\sigma}'}/W_a(\boldsymbol{\sigma})$. This construction yields direct access to channel-resolved activities (events per unit time) and provides a transparent separation of time scales between competing microscopic processes in the nonequilibrium steady state.

As a concrete demonstration, we apply the method to a stochastic lattice model for magnetization transport under directional drive and reactive (nonconserving) relaxation. We use the Ising model \cite{Ising1925} to describe the interaction energy between spins on a two-dimensional lattice through the Hamiltonian
\begin{equation}
\mathcal{H}=-J\sum_{\langle i,j\rangle}\sigma_i\sigma_j,
\label{eq:2}
\end{equation}
where $\sigma_i=\pm1$ and $\langle i,j\rangle$ denotes nearest-neighbor pairs. We consider an antiferromagnetic coupling and set $J=-1$, and implement two competing channels: (i) a drive-induced, magnetization-conserving exchange mechanism based on the Katz--Lebowitz--Spohn (KLS) dynamics~\cite{KLS1983,LeibowitzSpohn1984}; (ii) a thermal, nonconserving relaxation mechanism given by Glauber single-spin flips dynamic at temperature $T$~\cite{Glauber1963}.

The KLS dynamics~\cite{KLS1983,LeibowitzSpohn1984} is inspired by Kawasaki two-spin exchange~\cite{Kawasaki1966a,Kawasaki1966b}, but with exchanges biased by an external drive $E$ applied along the $+\hat{y}$ direction. The drive does not enter the Hamiltonian; instead, it contributes an external work term in the transition rates. Exchanges that transport an up spin ($+1$) along $+\hat{y}$ (equivalently, a down spin along $-\hat{y}$) are favored, while moves against the field are suppressed. This directional transport breaks detailed balance and maintains the system in a nonequilibrium steady state.

To encode the bias, we define $\phi_{ij}$ such that $\phi_{ij}=+1$ corresponds to an exchange that moves a $+1$ spin along $+\hat{y}$ (favored), $\phi_{ij}=-1$ to a move against the field (suppressed), and $\phi_{ij}=0$ to a transverse exchange. We adopt the following exchange rates for the KLS channel,
\begin{equation}
w_{ij\to ji} =
\frac{1-\delta_{\sigma_i,\sigma_j}}{1+\exp\!\left[\left(\Delta \mathcal{H}_{ij}-\phi_{ij} E\right)/k_BT\right]},
\label{eq:3}
\end{equation}
where the prefactor $1-\delta_{\sigma_i,\sigma_j}$ removes null moves (exchanges between equal spins), and $\Delta \mathcal{H}_{ij}$ is the energy difference associated with exchanging spins at sites $i$ and $j$, computed from Eq.~\eqref{eq:2}. The antiferromagnetic Ising model with KLS dynamics is well established in the literature, both in simulations and within pair approximations~\cite{PRL1989,PRA1990}, where one finds a reduction of $T_c$ with increasing field and qualitative changes in the nature of the transition, including the possibility of tricritical behavior separating continuous and discontinuous transitions. These features provide a useful baseline for comparison when the reactive channel is introduced.

The single-spin flip dynamics~\cite{Glauber1963} is used to model thermal reactive relaxation in contact with a heat bath at temperature $T$, and is defined by the rate
\begin{equation}
w_{i\to i^{\prime}} = \frac{1}{1+\exp\!\left(\Delta \mathcal{H}_{i}/k_BT\right)},
\label{eq:4}
\end{equation}
where $\Delta \mathcal{H}_{i}$ is the energy difference associated with flipping spin $\sigma_i$, computed from Eq.~\eqref{eq:2}, $k_B$ is the Boltzmann constant, and $T$ is the absolute temperature.

Within rejection-free CTKMC, the channel escape rates are 
\begin{equation}
W_G = \sum_i w_{i\to i^{\prime}}, \qquad
W_K = \sum_{\langle i,j\rangle} w_{ij\to ji},
\label{eq:5}
\end{equation}
and the total escape rate is $W = W_G + W_K$. Accordingly, the instantaneous probabilities that the next event is generated by the single-spin flip or KLS channel are respectively $q_G=W_G/W$ and $q_K=W_K/W$, with $q_G+q_K=1$.

A schematic representation of the framework applied to the present system is shown in Fig.~\ref{fig:1-1}. In this approach, the relative activity of the dynamical channels emerges self-consistently from the instantaneous configuration through their transition rates.

\begin{figure}
\begin{centering}
\begin{centering}
\includegraphics[scale=0.38]{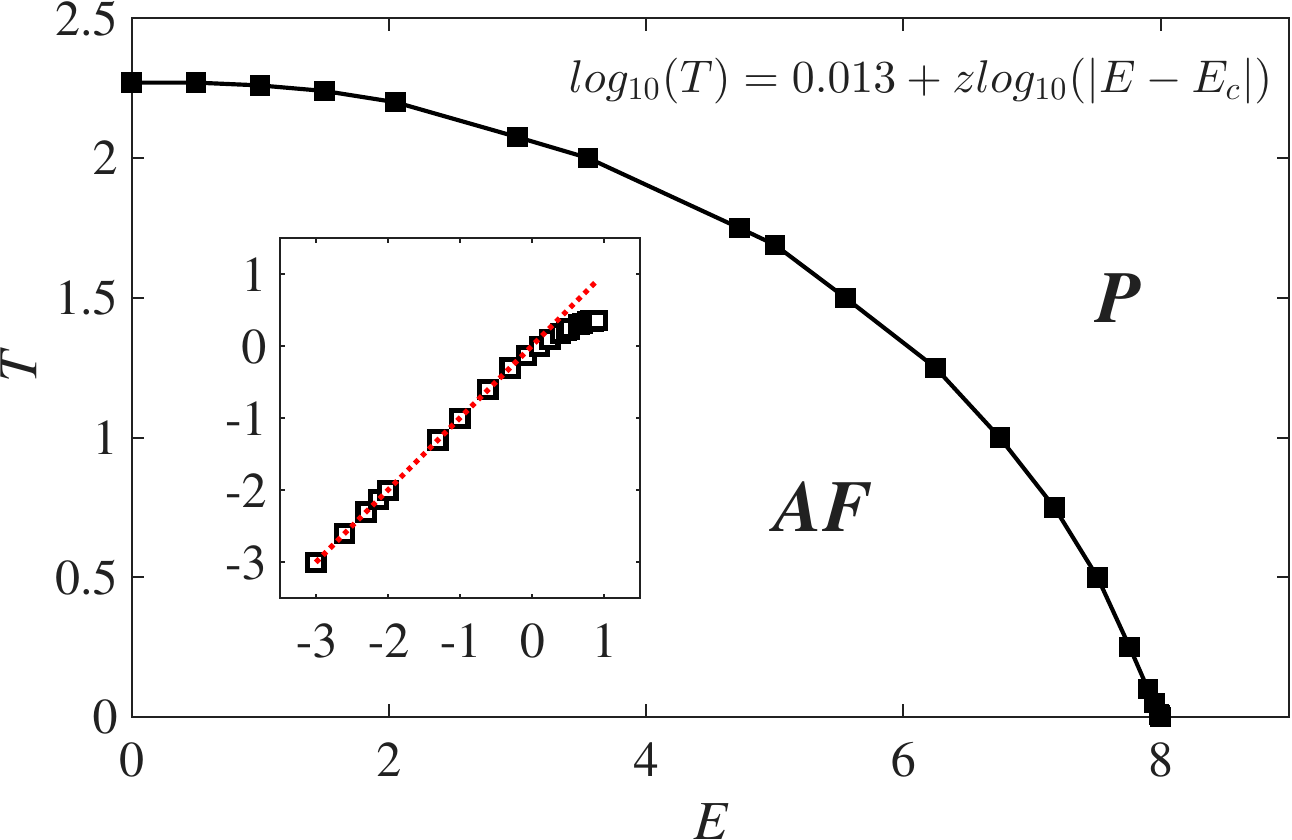}
\end{centering}
\caption{\textcolor{blue}{(Color online)} Phase diagram of the antiferromagnetic Ising model under competing dynamics between thermal relaxation via single-spin flips and driven diffusion induced by two-spin exchange (KLS) dynamics. The inset shows a logarithmic representation of the phase boundary, plotting $T$ as a function of $|E-E_c|$ with $E_c=8$. The linear behavior indicates a power-law scaling near the critical field in the limit $T\to0$, with exponent $z=1.001\pm0.001$. The fitted relation is displayed at the top of the panel. Error bars are smaller than the symbol size, and the lines are guides to the eye.}
\label{fig:1}
\end{centering}
\end{figure}

To evolve the system, we start from a random initial configuration and compute the instantaneous channel probabilities $q_G$ and $q_K$. A random number $r_1\in(0,1)$ is then drawn to select the active channel. If $r_1 \le q_G$, the single-spin flip dynamics is applied. The spin to be updated is selected by drawing a second random number $r_2\in(0,1)$, which determines the site through the cumulative sum corresponding to $r_2 W_G$. Otherwise, if $r_1 > q_G$, the KLS exchange dynamics is applied, and the pair of sites to be exchanged is determined from the cumulative sum corresponding to $r_2 W_K$. Each such event constitutes a single update step in the simulation and corresponds to a physical time increment determined by the total escape rate. In this formulation, configurations with larger escape rates $W$ produce shorter waiting times between successive events, naturally capturing the relative time scales of the competing processes. To ensure that the system reaches the stationary state, we perform $10^7$ update steps for thermalization, followed by an additional $10^7$ steps for measuring the observables. A total of $100$ independent realizations are generated for each set of control parameters $(T,E)$.

Within this framework, we investigate the phase diagram in the $T$–$E$ plane and compare the behavior of the present model, where two distinct dynamical channels act simultaneously, with the well-established case in which only the KLS dynamics is present~\cite{PRL1989,PRA1990}.

For $E=0$, the ground state of the system corresponds to antiferromagnetic ordering, where neighboring spins align antiparallel. We therefore use the staggered magnetization $m_L^{AF}$ as the order parameter, defined as
\begin{equation}
m_L^{AF} = \frac{1}{N}\sum_{i,j} (-1)^{i+j}\sigma_{ij},
\label{eq:6}
\end{equation}
where $i$ and $j$ denote the row and column indices of spin $\sigma_{ij}$ on the square lattice.

In the ordered phase and for small values of $E$, the most accessible process is the single-spin flip dynamics, since the associated energy change is typically smaller than that required for KLS exchanges. However, as the driving field increases, the energetic cost of exchanges becomes less relevant when spin transport occurs along the direction favored by the field. From these considerations, one expects the critical temperature to decrease with increasing $E$, since vertical spin exchanges tend to disrupt antiferromagnetic ordering. However, the presence of the nonconserving reactive channel introduces an additional mechanism that can locally restore antiferromagnetic alignment by flipping individual spins. As a result, this process can partially compensate for the disordering effect of the drive and stabilize magnetic ordering in regions of the $(T,E)$ plane where it would otherwise be suppressed.

This behavior is reflected in the phase diagram shown in Fig.~\ref{fig:1}. The diagram exhibits only continuous phase transitions over the entire range of parameters investigated, with temperatures spanning from $T=0.001$ up to the Néel temperature $T\approx2.267$~\cite{Onsager1944}. As expected, the antiferromagnetic ordering decreases as the driving field $E$ increases. However, this decay is significantly weaker than that observed in the model with pure KLS dynamics, and no tricritical point separating first- and second-order transitions is detected at low temperatures. An additional feature of the phase boundary is a power-law behavior near the critical field in the limit $T\to0$, characterized by an exponent close to $1$. A linear fit of the phase boundary in logarithmic scale is shown in the inset of Fig.~\ref{fig:1}, together with the corresponding fitting equation.

\begin{figure}
\begin{centering}
\begin{centering}
\vspace{0.1cm}
\includegraphics[scale=0.22]{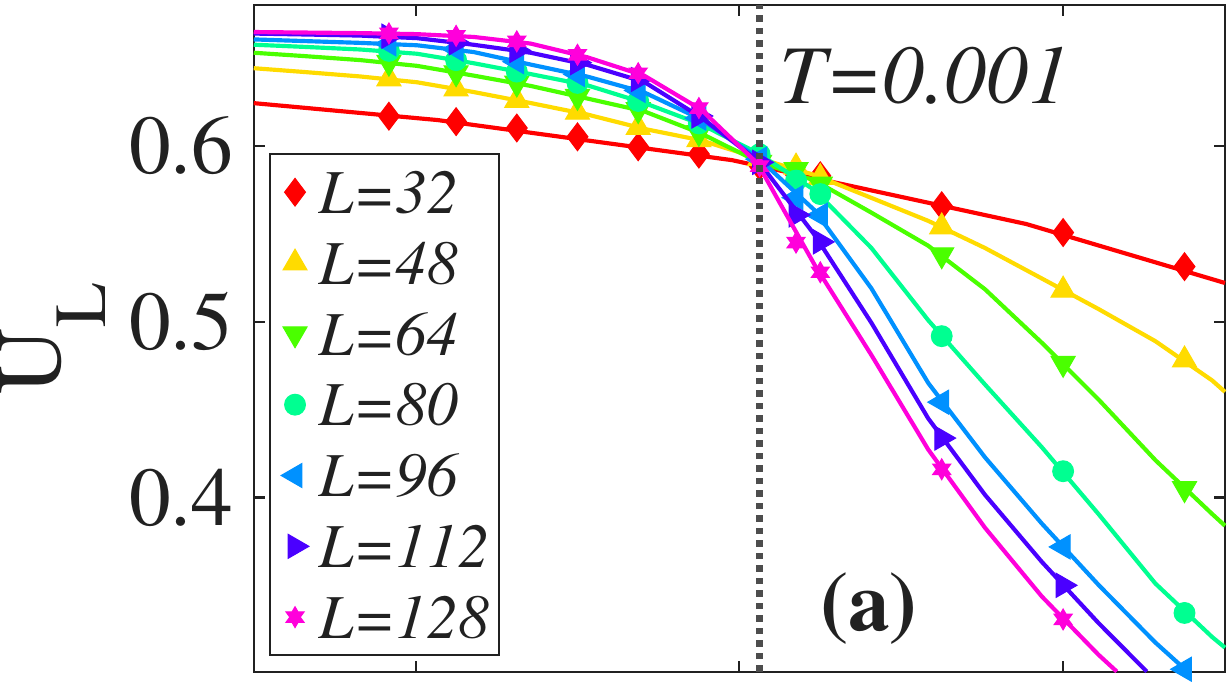} \includegraphics[scale=0.22]{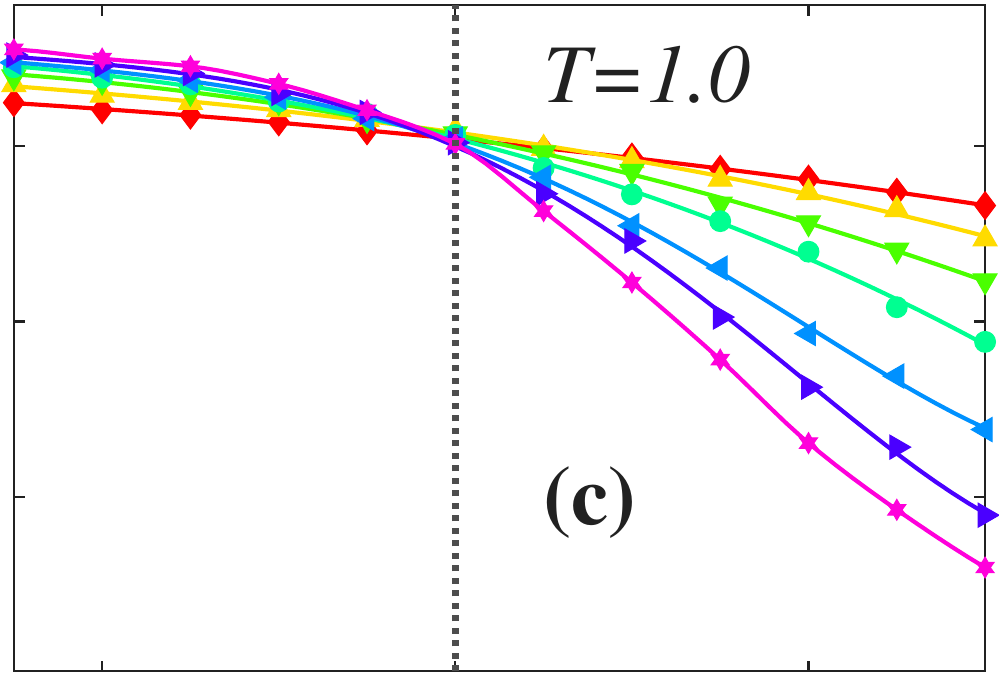}
\includegraphics[scale=0.22]{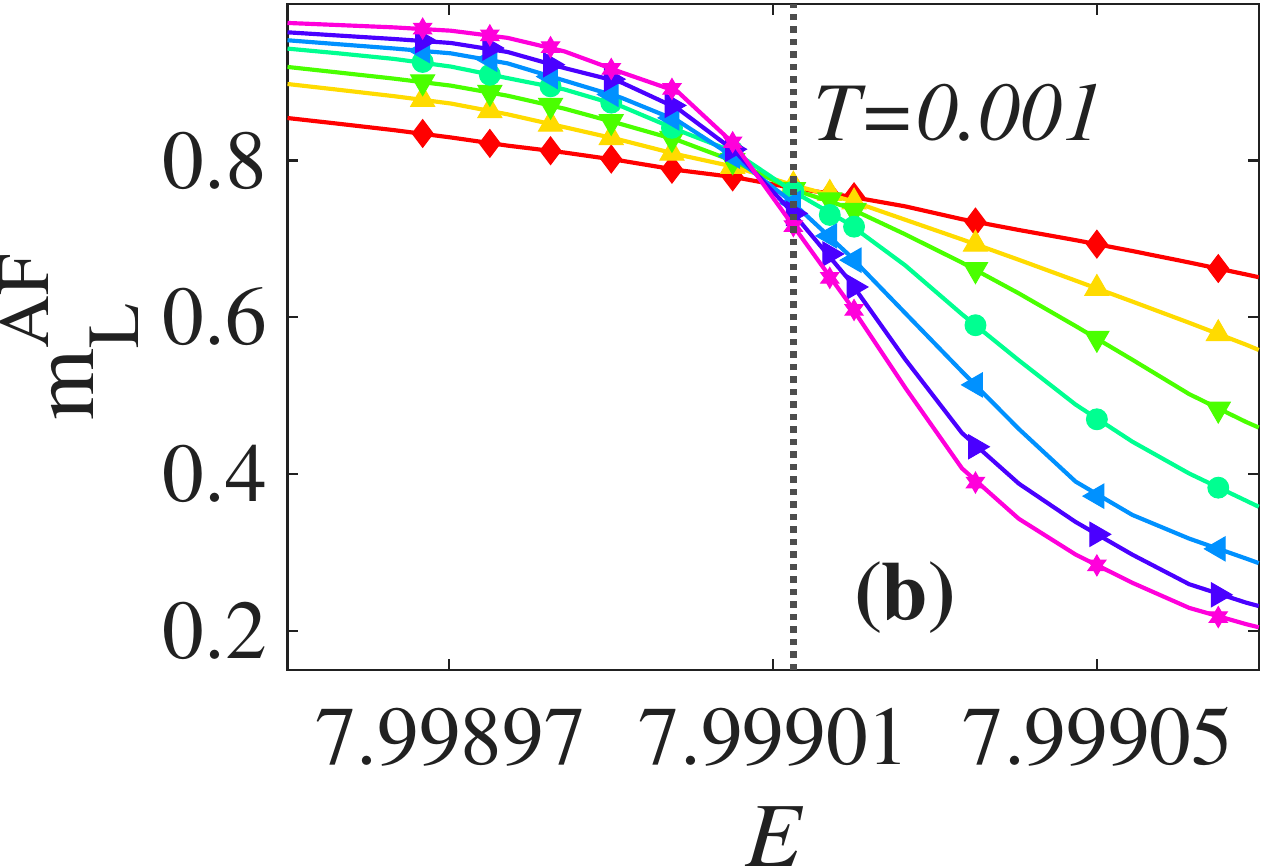} \includegraphics[scale=0.22]{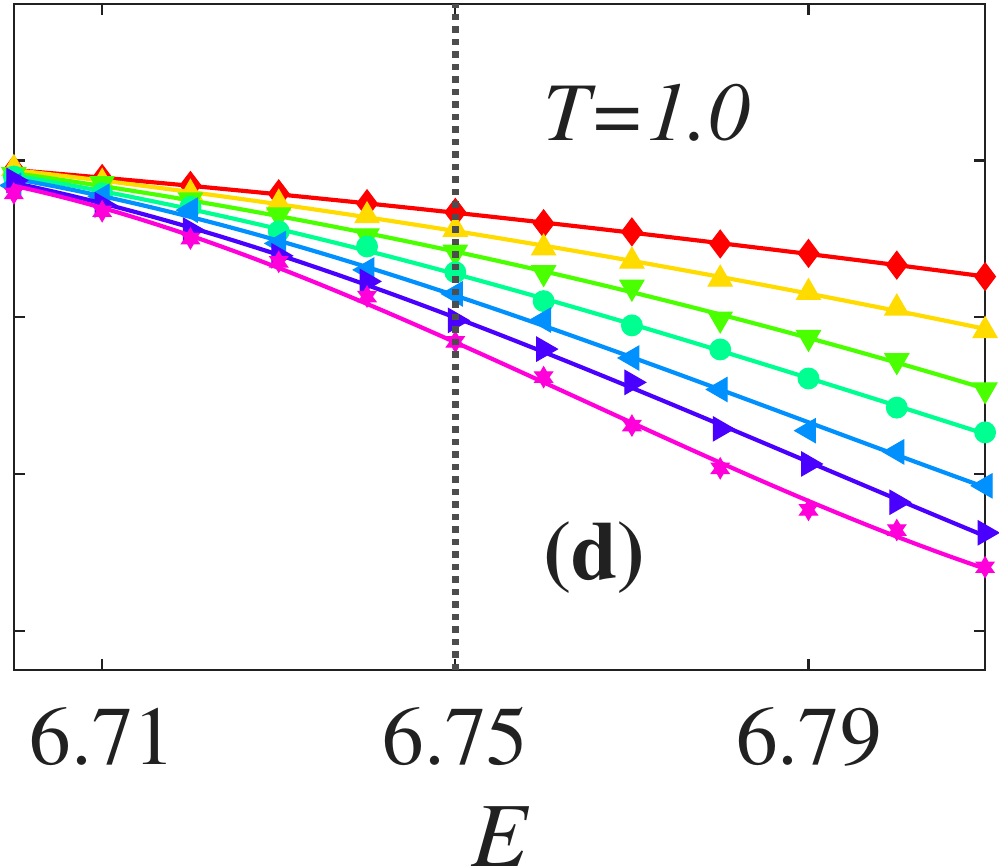}
\end{centering}
\caption{\textcolor{blue}{(Color online)} Comparison between the different critical regimes observed in the system through the analysis of $m_L^{AF}$ and the Binder cumulant $U_L$. Panels (a) and (b) show $U_L$ and $m_L^{AF}$ at $T = 0.001$ as functions of $E$, respectively, while panels (c) and (d) present $U_L$ and $m_L^{AF}$ at $T = 1$. In all panels, results are shown for different system sizes, as shown in panel (a). The vertical black dotted lines indicate the critical point of the phase transition listed in Table~\ref{Tab.1}, determined from the crossing of the $U_L$ curves. Error bars are smaller than the symbol size, and the lines are guides to the eye.}
\label{fig:2}
\end{centering}
\end{figure}

The order of the phase transition was determined from the behavior of the order parameter and its probability distribution (see Fig. 1 in the Supplemental Material). No evidence of phase coexistence was observed: the probability distribution of the order parameter does not exhibit multiple peaks, and no hysteresis loops were detected in the simulations. 

To further characterize the transition, we also analyze the Binder cumulant for different system sizes, defined as~\cite{binder2021monte,binder1981}
\begin{equation}
U_L = 1 - \frac{\langle (m_L^{AF})^{4} \rangle}{3 \langle (m_L^{AF})^{2} \rangle^{2}} .
\label{eq:7}
\end{equation}

Besides providing a useful diagnostic of the order of the transition, the Binder cumulant allows for a precise determination of the critical point through the crossing of curves corresponding to different lattice sizes~\cite{tsai}. 

A clear illustration of the critical behavior at low and intermediate temperatures is shown in Fig.~\ref{fig:2}. In Figs.~\ref{fig:2}(a) and~\ref{fig:2}(b), we present $U_L$ and $m_L^{AF}$ for $T=0.001$, respectively, while Figs.~\ref{fig:2}(c) and~\ref{fig:2}(d) display the same quantities for $T=1.0$. In both cases, the critical point can be clearly identified from the crossing of the Binder cumulant curves. However, an important difference emerges in the behavior of the order parameter. At $T=0.001$, the curves of $m_L^{AF}$ for different system sizes collapse much more closely near the critical point than those obtained at $T=1.0$. This indicates a significantly smaller critical exponent for the order parameter in the low-temperature regime.

\begin{table}
\caption{Critical field $E_c$ for different temperatures $T$, together with estimates of the correlation-length exponent $1/\nu$ and the order-parameter exponent ratio $\beta/\nu$.}

\begin{onehalfspace}
\centering
\begin{tabular*}{\columnwidth}{@{\extracolsep{\fill}}cccc}

\hline 
$T $ & $E_c$ & $1/\nu$ & $\beta/\nu$
\tabularnewline
\hline 

$0.001$  & $7.9990125\pm0.00001$  & $1.37\pm0.04$  & $0.021\pm0.01$
\tabularnewline

$0.01$  & $7.9901625\pm0.0001$  & $1.40\pm0.04$  & $0.029\pm0.02$
\tabularnewline

$0.1$  & $7.90155\pm0.001$  & $1.40\pm0.03$  & $0.024\pm0.02$
\tabularnewline

$1.0$  & $6.75\pm0.01$  & $1.04\pm0.05$  & $0.122\pm0.02$
\tabularnewline

\hline 

\end{tabular*}
\end{onehalfspace}
\label{Tab.1}
\end{table}

To estimate the critical exponents, we employ the standard finite-size scaling relations valid near the critical point~\cite{binder2021monte},
\begin{subequations}
\begin{align}
m_L &= L^{-\beta/\nu} m_{0}(L^{1/\nu}\epsilon), \label{eq:8a}\\
U_L^{\prime} &= L^{1/\nu} \mathcal{U}_{0}^{\prime}(L^{1/\nu}\epsilon), \label{eq:8b}
\end{align}
\end{subequations}
where $\epsilon=(E-E_c)/E_c$ is the reduced field, and $\beta$ and $\nu$ are the critical exponents associated with the order parameter and the correlation length, respectively.

The exponent ratios were obtained from linear fits of the relevant quantities as functions of the system size in logarithmic scale. The resulting estimates, together with the corresponding critical points obtained from the Binder cumulant crossings, are reported in Table~\ref{Tab.1}. These results indicate that in the low-temperature region of the phase diagram ($T\to0$), the exponent associated with the order parameter approaches zero. In contrast, for higher temperatures, such as $T=1.0$, the system recovers the universality class of the two-dimensional Ising model~\cite{odor2004}.

\begin{figure}
\begin{centering}
\begin{centering}
\includegraphics[scale=0.22]{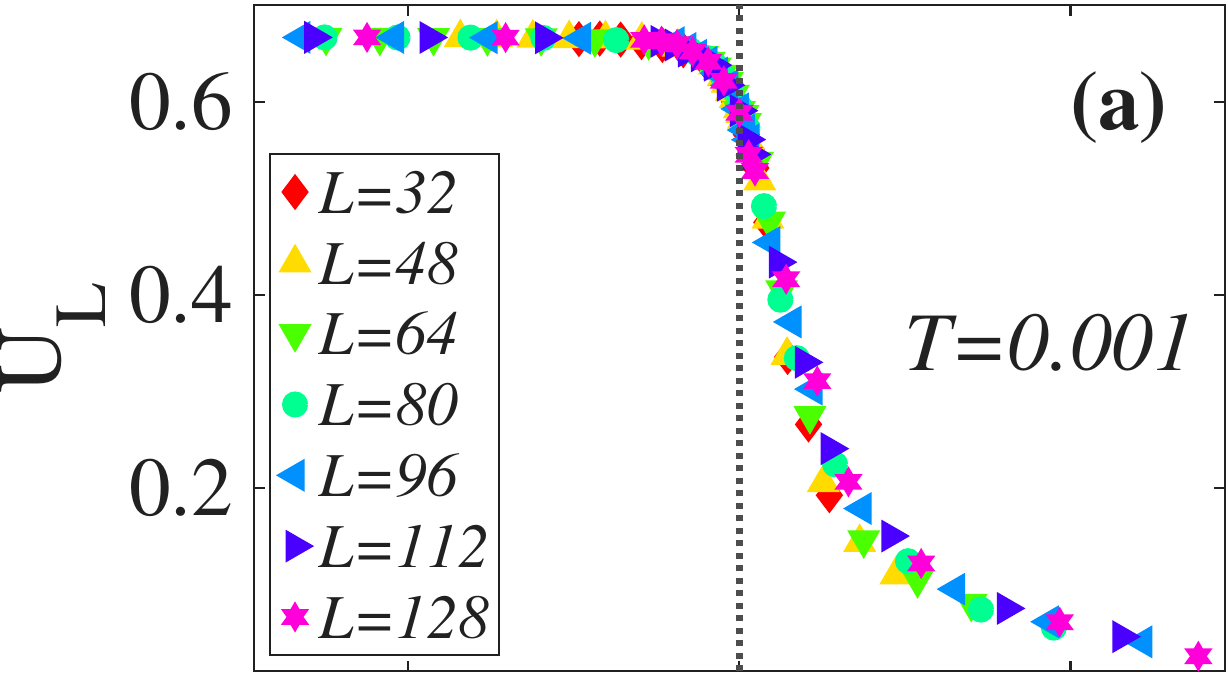} \includegraphics[scale=0.22]{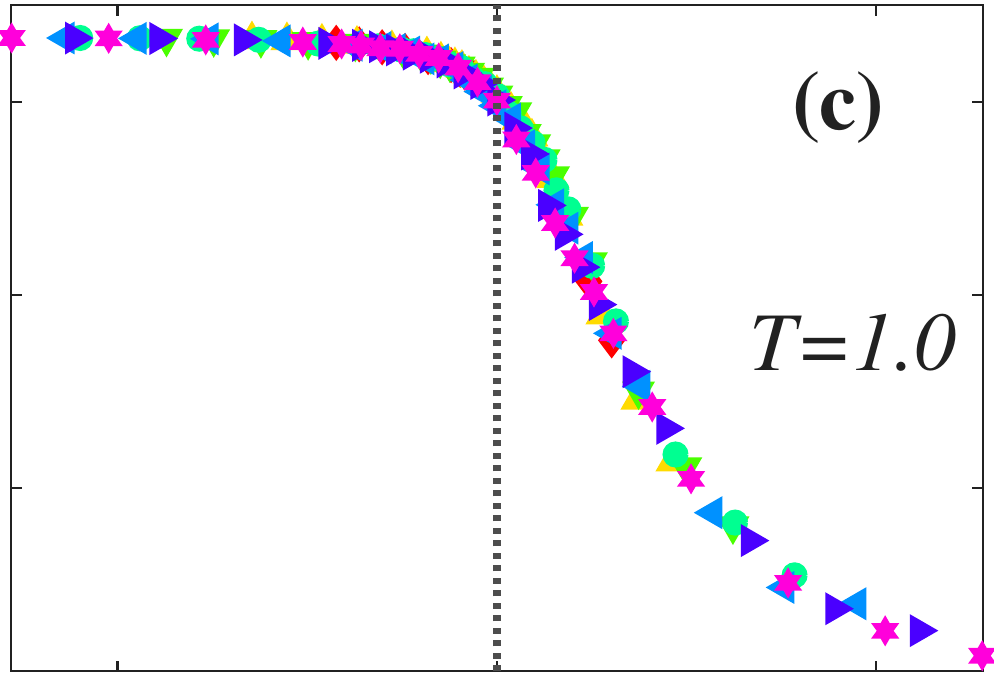}
\includegraphics[scale=0.22]{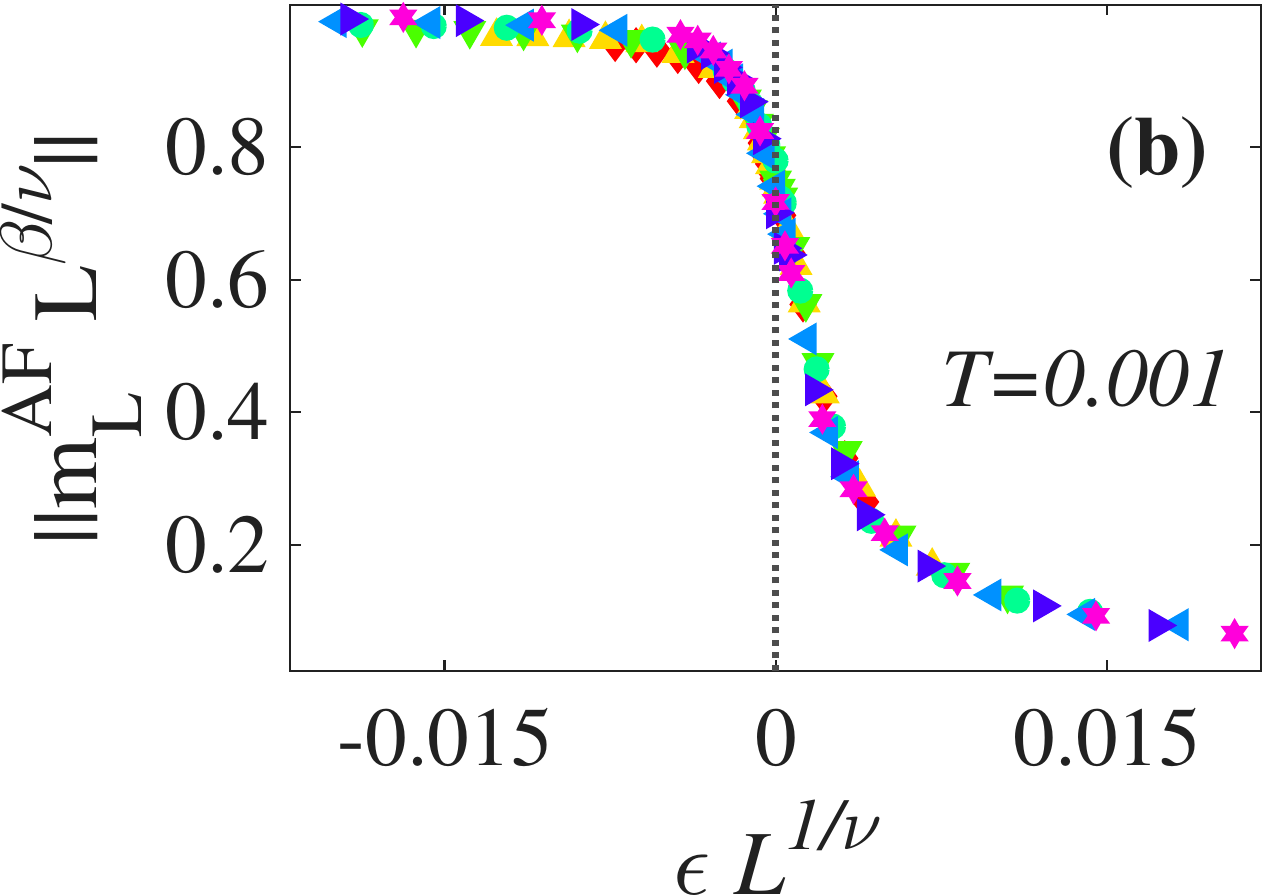} \includegraphics[scale=0.22]{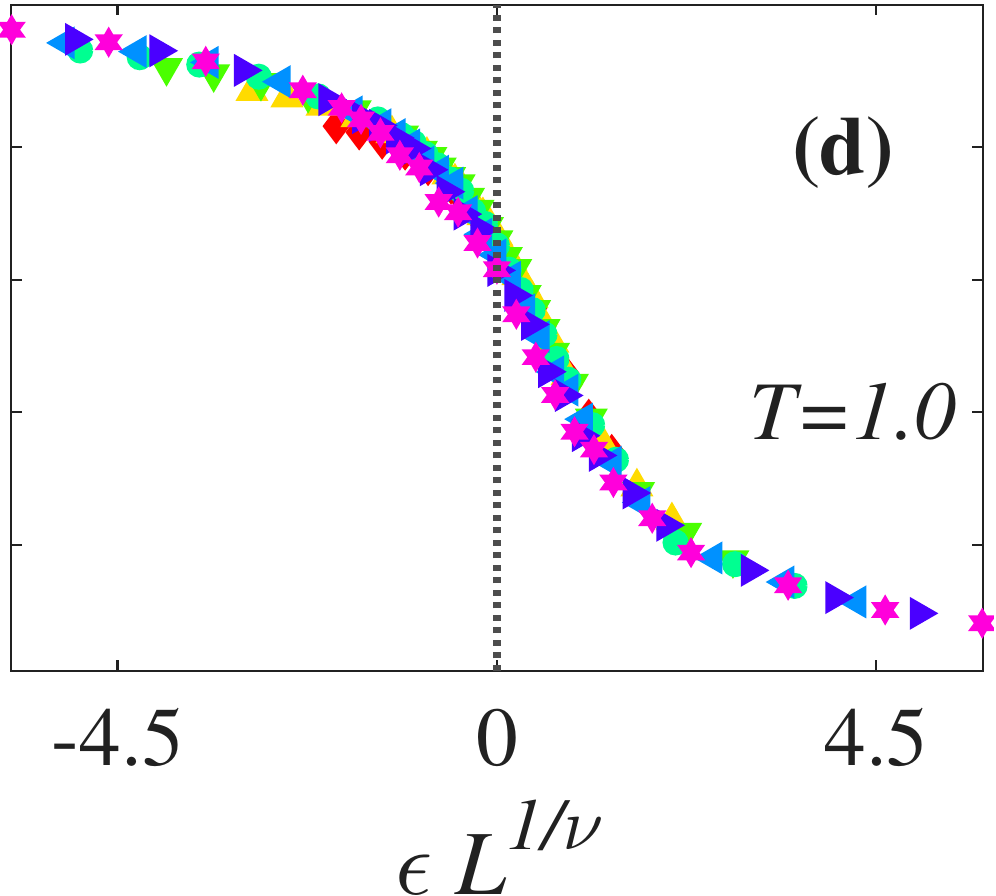}
\end{centering}
\caption{\textcolor{blue}{(Color online)} Data collapse of the curves shown in Fig.~\ref{fig:2}, yielding the scaling functions of Eqs.~\eqref{eq:8a} and~\eqref{eq:8b} using the critical exponents listed in Table~\ref{Tab.1}. For clarity, the $m_0$ curves were normalized by $m_0 = 1.1322$ in panel (b) and $m_0 = 1.67$ in panel (d), without affecting the quality of the collapse.}
\label{fig:3}
\end{centering}
\end{figure}

To further verify the consistency of the estimated critical exponents, we performed a data collapse to obtain the scaling functions defined in Eqs.~\eqref{eq:8a} and~\eqref{eq:8b}. For $T=0.001$, the resulting collapses yielding the scaling functions $m_{0}$ and $\mathcal{U}_{0}^{\prime}$ are shown in Figs.~\ref{fig:3}(a) and~\ref{fig:3}(b), respectively. For comparison, the corresponding collapses for $T=1.0$ are presented in Figs.~\ref{fig:3}(c) and~\ref{fig:3}(d). The data collapse was obtained using the exponent ratios listed in Table \ref{Tab.1}. The quality of the collapse provides strong support for the exponent estimates and for the identification of the universality class in the corresponding parameter regimes.

\begin{figure}
\begin{centering}
\begin{centering}
\includegraphics[scale=0.40]{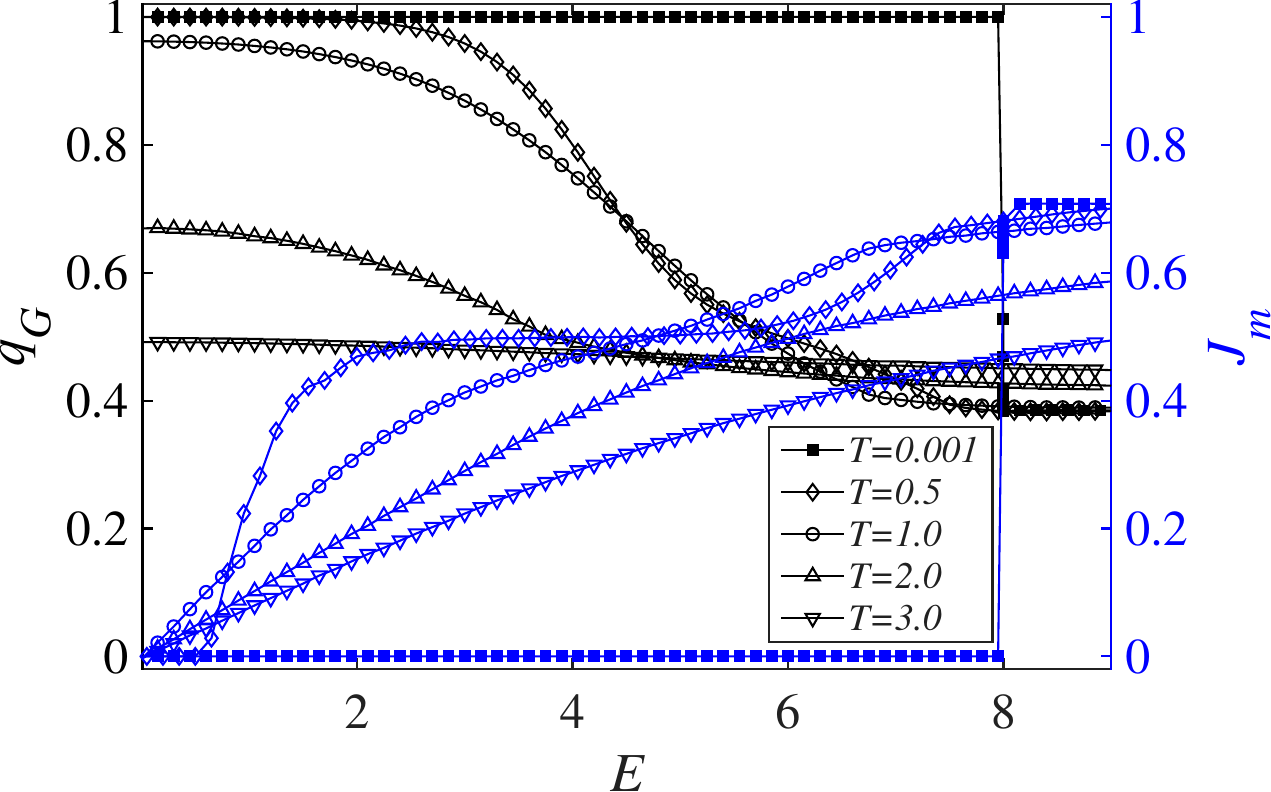}
\end{centering}
\caption{\textcolor{blue}{(Color online)} Stationary values of the single-spin-flip activation rate, $q_G$, and of the magnetization current, $J_m$, as functions of $E$ for different values of $T$.}
\label{fig:4}
\end{centering}
\end{figure}

An additional feature of the rejection-free CTKMC framework is the ability to directly measure the activity of each dynamical channel. Since the transition rates depend on the instantaneous configuration of the system, the corresponding channel probabilities also vary with the system parameters. Examples of this behavior are shown in Fig.~\ref{fig:4}, where we present the dependence of $q_G$ (and consequently $q_K=1-q_G$). For low values of $T$ and $E$, the dominant mechanism is the single-spin flip dynamics, as discussed earlier in this Letter. However, as the disorder parameters increase, the KLS dynamics becomes progressively more active, leading to the emergence of a magnetization current $J_m$.

The magnetization current is measured only during KLS updates and is defined as the difference between spin exchanges favored by the field and those occurring against the field, normalized by the total number of exchange events. The corresponding curves for $J_m$ are also shown in Fig.~\ref{fig:4}, where a clear increase of the current with the driving field is observed. Both quantities eventually reach saturation values, approximately $J_m \approx 0.708$ and $q_G \approx 0.385$. The curves shown correspond to system size $L=128$, but no significant size dependence was observed for the range of system sizes investigated ($32 \le L \le 128$).

Another relevant quantity in the present context is the dynamic exponent $\xi$, associated with the growth of the average domain size during the evolution from a disordered initial state toward the stationary regime~\cite{sadiq1983,mendes1992}. For dynamics that do not conserve the order parameter, such as the single-spin flip dynamics, one expects $\xi=1/2$, whereas for dynamics that conserve the order parameter, the expected value is $\xi=1/3$. For the temperatures listed in Table~\ref{Tab.1}, we measured $\xi$ at three representative values of the driving field: $E=1$ (below the critical field), $E\approx E_c$, and $E=10$ (above the critical field). The estimated values indicate a clear crossover between distinct regimes, with $\xi \approx 0.47$ for $E=1$, $\xi \approx 0.38$ near $E_c$, and $\xi \approx 0.29$ for $E=10$. No significant dependence on system size or temperature was detected within the range investigated (see Fig. 2 in the Supplemental Material).

In summary, we introduced a rejection-free continuous-time kinetic Monte Carlo framework to investigate stochastic systems governed by multiple competing dynamical channels. Applying this approach to the driven antiferromagnetic KLS model with an additional Glauber relaxation mechanism, we showed that the coexistence of conservative and nonconservative processes substantially modifies the nonequilibrium critical behavior. In particular, the ordered phase becomes significantly stabilized against the external drive $E$, and the phase diagram exhibits a power-law boundary of the form $T\sim|E-E_c|$ when $T\to0$, together with purely continuous transitions and a modification of the effective critical behavior in the low-temperature limit. Beyond the specific example studied here, the proposed framework provides a general strategy for studying systems with concurrent dynamical mechanisms. In contrast with many nonequilibrium models where the relative activity of different processes is imposed externally, here it emerges naturally from the microscopic transition rates, removing the need to prescribe arbitrary probabilities. This makes the approach particularly suitable for systems where conservative and reactive processes coexist, such as reaction–diffusion systems~\cite{reacdif1,odor2004}, driven lattice gases~\cite{LeibowitzSpohn1984,driven1}, active matter~\cite{active1,active2}, surface growth models~\cite{surface1}, and systems coupled to multiple reservoirs~\cite{ts21,ts22}. More broadly, the framework opens the possibility of exploring how competing dynamical channels shape collective behavior and critical phenomena in complex nonequilibrium systems, including extensions to multiplex structures where different layers host distinct dynamical processes that interact through state-dependent rates~\cite{mult1,mult2}.

R. A. Dumer was supported by the Coordenação de Aperfeiçoamento de Pessoal de Nível Superior - Brasil (CAPES) - Finance Code 001, and by a PhD Scholarship (CNPq) No. $140141/2024-3$. This work was developed within the scope of the project i3N under the project UID/50025/2025 and the Associate Laboratory i3N – LA/P/0037/2020, financed by the FCT. 

\bibliography{references.bib}

@article{Glauber1963,
  author  = {Glauber, Roy J.},
  title   = {Time-Dependent Statistics of the Ising Model},
  journal = {J. Math. Phys.},
  volume  = {4},
  number  = {2},
  pages   = {294--307},
  year    = {1963},
  doi     = {10.1063/1.1703954},
  url     = {https://doi.org/10.1063/1.1703954}
}

@article{Kawasaki1966a,
  author  = {Kawasaki, Kyozi},
  title   = {Diffusion Constants near the Critical Point for Time-Dependent Ising Models. I},
  journal = {Phys. Rev.},
  volume  = {145},
  number  = {1},
  pages   = {224--230},
  year    = {1966},
  doi     = {10.1103/PhysRev.145.224},
  url     = {https://link.aps.org/doi/10.1103/PhysRev.145.224}
}

@article{Kawasaki1966b,
  author  = {Kawasaki, Kyozi},
  title   = {Diffusion Constants near the Critical Point for Time-Dependent Ising Models. II},
  journal = {Phys. Rev.},
  volume  = {148},
  number  = {1},
  pages   = {375--381},
  year    = {1966},
  doi     = {10.1103/PhysRev.148.375},
  url     = {https://link.aps.org/doi/10.1103/PhysRev.148.375}
}

@article{PRA1989,
  author  = {Tom\'e, T\^ania and de Oliveira, M\'ario J.},
  title   = {Self-organization in a kinetic Ising model},
  journal = {Phys. Rev. A},
  volume  = {40},
  number  = {11},
  pages   = {6643--6646},
  year    = {1989},
  doi     = {10.1103/PhysRevA.40.6643},
  url     = {https://link.aps.org/doi/10.1103/PhysRevA.40.6643}
}

@article{PRE2000,
  author  = {Szolnoki, Attila},
  title   = {Phase transitions in the kinetic Ising model with competing dynamics},
  journal = {Phys. Rev. E},
  volume  = {62},
  number  = {5},
  pages   = {7466--7469},
  year    = {2000},
  doi     = {10.1103/PhysRevE.62.7466},
  url     = {https://link.aps.org/doi/10.1103/PhysRevE.62.7466}
}

@article{PRE1999,
  author  = {Le\~ao, J. R. S. and Grandi, B. C. S. and Figueiredo, W.},
  title   = {Competitive dynamics in a three-dimensional Ising model},
  journal = {Phys. Rev. E},
  volume  = {60},
  number  = {5},
  pages   = {5367--5370},
  year    = {1999},
  doi     = {10.1103/PhysRevE.60.5367},
  url     = {https://link.aps.org/doi/10.1103/PhysRevE.60.5367}
}

@article{PRE2025,
  author  = {Dumer, R. A. and da Costa, D. R. and Godoy, M.},
  title   = {Classical three-dimensional Heisenberg model with competing dynamics},
  journal = {Phys. Rev. E},
  volume  = {112},
  number  = {4},
  pages   = {044119},
  year    = {2025},
  doi     = {10.1103/PhysRevE.112.044119},
  url     = {https://link.aps.org/doi/10.1103/PhysRevE.112.044119}
}

@article{FrontChem2019,
  author  = {Andersen, Mie and Panosetti, Chiara and Reuter, Karsten},
  title   = {A Practical Guide to Surface Kinetic Monte Carlo Simulations},
  journal = {Front. Chem.},
  volume  = {7},
  pages   = {202},
  year    = {2019},
  doi     = {10.3389/fchem.2019.00202},
  url     = {https://www.frontiersin.org/journals/chemistry/articles/10.3389/fchem.2019.00202}
}

@article{Gillespie1977,
  author  = {Gillespie, Daniel T.},
  title   = {Exact stochastic simulation of coupled chemical reactions},
  journal = {J. Phys. Chem.},
  volume  = {81},
  number  = {25},
  pages   = {2340--2361},
  year    = {1977},
  doi     = {10.1021/j100540a008},
  url     = {https://doi.org/10.1021/j100540a008}
}

@article{BKL1975,
  author  = {Bortz, A. B. and Kalos, M. H. and Lebowitz, J. L.},
  title   = {A new algorithm for Monte Carlo simulation of Ising spin systems},
  journal = {J. Comput. Phys.},
  volume  = {17},
  number  = {1},
  pages   = {10--18},
  year    = {1975},
  doi     = {10.1016/0021-9991(75)90060-1},
  url     = {https://www.sciencedirect.com/science/article/pii/0021999175900601}
}

@article{LeibowitzSpohn1984,
  author  = {Katz, Sheldon and Lebowitz, Joel L. and Spohn, Herbert},
  title   = {Nonequilibrium steady states of stochastic lattice gas models of fast ionic conductors},
  journal = {J. Stat. Phys.},
  volume  = {34},
  number  = {3},
  pages   = {497--537},
  year    = {1984},
  doi     = {10.1007/BF01018556},
  url     = {https://doi.org/10.1007/BF01018556}
}

@article{KLS1983,
  author  = {Katz, Sheldon and Lebowitz, Joel L. and Spohn, H.},
  title   = {Phase transitions in stationary nonequilibrium states of model lattice systems},
  journal = {Phys. Rev. B},
  volume  = {28},
  number  = {3},
  pages   = {1655--1658},
  year    = {1983},
  doi     = {10.1103/PhysRevB.28.1655},
  url     = {https://link.aps.org/doi/10.1103/PhysRevB.28.1655}
}

@article{PRL1989,
  author  = {Leung, K.-t. and Schmittmann, B. and Zia, R. K. P.},
  title   = {Phase transitions in a driven lattice gas with repulsive interactions},
  journal = {Phys. Rev. Lett.},
  volume  = {62},
  number  = {15},
  pages   = {1772--1775},
  year    = {1989},
  doi     = {10.1103/PhysRevLett.62.1772},
  url     = {https://link.aps.org/doi/10.1103/PhysRevLett.62.1772}
}

@article{PRA1990,
  author  = {Dickman, Ronald},
  title   = {Driven lattice gas with repulsive interactions: Mean-field theory},
  journal = {Phys. Rev. A},
  volume  = {41},
  number  = {4},
  pages   = {2192--2195},
  year    = {1990},
  doi     = {10.1103/PhysRevA.41.2192},
  url     = {https://link.aps.org/doi/10.1103/PhysRevA.41.2192}
}

@article{ShockleyRead1952,
  title = {Statistics of the Recombinations of Holes and Electrons},
  author = {Shockley, W. and Read, W. T.},
  journal = {Phys. Rev.},
  volume = {87},
  issue = {5},
  pages = {835--842},
  numpages = {0},
  year = {1952},
  month = {Sep},
  publisher = {American Physical Society},
  doi = {10.1103/PhysRev.87.835},
  url = {https://link.aps.org/doi/10.1103/PhysRev.87.835}
}

@article{Hall1952,
  title = {Electron-Hole Recombination in Germanium},
  author = {Hall, R. N.},
  journal = {Phys. Rev.},
  volume = {87},
  issue = {2},
  pages = {387--387},
  numpages = {0},
  year = {1952},
  month = {Jul},
  publisher = {American Physical Society},
  doi = {10.1103/PhysRev.87.387},
  url = {https://link.aps.org/doi/10.1103/PhysRev.87.387}
}

@article{DoyleFullerNewman1993,
  author  = {Doyle, Marc and Fuller, Thomas F. and Newman, John},
  title   = {Modeling of Galvanostatic Charge and Discharge of the Lithium/Polymer/Insertion Cell},
  journal = {J. Electrochem. Soc.},
  volume  = {140},
  pages   = {1526--1533},
  year    = {1993},
  doi     = {10.1149/1.2221597}
}

@article{Bazant2013,
  author  = {Bazant, Martin Z.},
  title   = {Theory of Chemical Kinetics and Charge Transfer based on Nonequilibrium Thermodynamics},
  journal = {Acc. Chem. Res.},
  volume  = {46},
  pages   = {1144--1160},
  year    = {2013},
  doi     = {10.1021/ar300145c}
}

@article{Tserkovnyak2002,
  title = {Enhanced Gilbert Damping in Thin Ferromagnetic Films},
  author = {Tserkovnyak, Yaroslav and Brataas, Arne and Bauer, Gerrit E. W.},
  journal = {Phys. Rev. Lett.},
  volume = {88},
  issue = {11},
  pages = {117601},
  numpages = {4},
  year = {2002},
  month = {Feb},
  publisher = {American Physical Society},
  doi = {10.1103/PhysRevLett.88.117601},
  url = {https://link.aps.org/doi/10.1103/PhysRevLett.88.117601}
}

@article{Chumak2015,
  author  = {Chumak, Andrii V. and Vasyuchka, Vitaliy I. and Serga, Alexey A. and Hillebrands, Burkard},
  title   = {Magnon spintronics},
  journal = {Nat. Phys.},
  volume  = {11},
  pages   = {453--461},
  year    = {2015},
  doi     = {10.1038/nphys3347}
}

@article{Cornelissen2015,
  author  = {Cornelissen, Lars J. and Liu, Jing and Duine, Rembert A. and Youssef, Joseph Ben and van Wees, Bart J.},
  title   = {Long-distance transport of magnon spin information in a magnetic insulator at room temperature},
  journal = {Nat. Phys.},
  volume  = {11},
  pages   = {1022--1026},
  year    = {2015},
  doi     = {10.1038/nphys3465}
}

@article{HohenbergHalperin1977,
  title = {Theory of dynamic critical phenomena},
  author = {Hohenberg, P. C. and Halperin, B. I.},
  journal = {Rev. Mod. Phys.},
  volume = {49},
  issue = {3},
  pages = {435--479},
  numpages = {0},
  year = {1977},
  month = {Jul},
  publisher = {American Physical Society},
  doi = {10.1103/RevModPhys.49.435},
  url = {https://link.aps.org/doi/10.1103/RevModPhys.49.435}
}

@article{Parmeggiani2003,
  title = {Phase Coexistence in Driven One-Dimensional Transport},
  author = {Parmeggiani, A. and Franosch, T. and Frey, E.},
  journal = {Phys. Rev. Lett.},
  volume = {90},
  issue = {8},
  pages = {086601},
  numpages = {4},
  year = {2003},
  month = {Feb},
  publisher = {American Physical Society},
  doi = {10.1103/PhysRevLett.90.086601},
  url = {https://link.aps.org/doi/10.1103/PhysRevLett.90.086601}
}

@article{Evans2005,
  title = {Shock formation in an exclusion process with creation and annihilation},
  author = {Evans, M. R. and Juh\'asz, R. and Santen, L.},
  journal = {Phys. Rev. E},
  volume = {68},
  issue = {2},
  pages = {026117},
  numpages = {8},
  year = {2003},
  month = {Aug},
  publisher = {American Physical Society},
  doi = {10.1103/PhysRevE.68.026117},
  url = {https://link.aps.org/doi/10.1103/PhysRevE.68.026117}
}

@article{Ising1925,
  title = {Beitrag zur Theorie des Ferromagnetismus},
  author = {Ising, Ernst},
  journal = {Zeitschrift für Physik},
  volume = {31},
  issue = {1},
  pages = {253},
  numpages = {5},
  year = {1925},
  month = {Aug},
  doi = {10.1007/BF02980577},
  url = {https://doi.org/10.1007/BF02980577}
}

@article{Onsager1944,
  title = {Crystal Statistics. I. A Two-Dimensional Model with an Order-Disorder Transition},
  author = {Onsager, Lars},
  journal = {Phys. Rev.},
  volume = {65},
  issue = {3-4},
  pages = {117--149},
  numpages = {0},
  year = {1944},
  month = {Feb},
  publisher = {American Physical Society},
  doi = {10.1103/PhysRev.65.117},
  url = {https://link.aps.org/doi/10.1103/PhysRev.65.117}
}

@book{binder2021monte,
  title={Monte Carlo Methods in Statistical Physics},
  author={Newman, M. E. J. and Barkema, G. T.},
  year={1999},
  publisher={Oxford University Press, NewYork, US},
  url = {https://link.springer.com/book/10.1007/978-3-642-82803-4}
}

@article{binder1981,
  title = {Finite size scaling analysis of ising model block distribution functions},
  author = {Binder, K.},
  journal = {Zeitschrift für Physik B Condensed Matter},
  volume = {43},
  issue = {2},
  pages = {119--140},
  numpages = {21},
  year = {1981},
  doi = {10.1007/BF01293604},
  url = {https://doi.org/10.1007/BF01293604}
}

@Article{tsai,
	title={{Fourth-Order Cumulants to Characterize the Phase Transitions of a Spin-1 Ising Model}},
	author={Shan-Ho Tsai and S. R. Salinas },
	journal={Brazilian Journal of Physics},
	volume={28},
    issue = {1},
	pages={58},
	year={1998},
	url={https://doi.org/10.1590/S0103-97331998000100008},
}

@article{odor2004,
  title = {Universality classes in nonequilibrium lattice systems},
  author = {\'Odor, G\'eza},
  journal = {Rev. Mod. Phys.},
  volume = {76},
  issue = {3},
  pages = {663--724},
  numpages = {0},
  year = {2004},
  month = {Aug},
  publisher = {American Physical Society},
  doi = {10.1103/RevModPhys.76.663},
  url = {https://link.aps.org/doi/10.1103/RevModPhys.76.663}
}

@article{mendes1992,
doi = {10.1088/0305-4470/25/1/012},
url = {https://doi.org/10.1088/0305-4470/25/1/012},
year = {1992},
month = {jan},
publisher = {},
volume = {25},
number = {1},
pages = {73},
author = {J F F Mendes and S Cornell and M Droz and E J S Lage},
title = {Nonuniversal critical behaviour in the 1D BEG model with Kawasaki dynamics},
journal = {Journal of Physics A: Mathematical and General},
}

@article{sadiq1983,
  title = {Kinetics of Domain Growth in Two Dimensions},
  author = {Sadiq, A. and Binder, K.},
  journal = {Phys. Rev. Lett.},
  volume = {51},
  issue = {8},
  pages = {674--677},
  numpages = {0},
  year = {1983},
  month = {Aug},
  publisher = {American Physical Society},
  doi = {10.1103/PhysRevLett.51.674},
  url = {https://link.aps.org/doi/10.1103/PhysRevLett.51.674}
}

@article{arek2025,
  title = {When Does Population Diversity Matter? A Unified Framework for Binary-Choice Dynamics},
  author = {J\ifmmode \mbox{\k{e}}\else \k{e}\fi{}drzejewski, Arkadiusz and Mendes, Jos\'e F. F.},
  journal = {Phys. Rev. Lett.},
  volume = {135},
  issue = {21},
  pages = {217401},
  numpages = {9},
  year = {2025},
  month = {Nov},
  publisher = {American Physical Society},
  doi = {10.1103/4db2-7dpd},
  url = {https://link.aps.org/doi/10.1103/4db2-7dpd}
}

@article{reacdif1,
  title = {Non-equilibrium critical
phenomena and phase transitions into absorbing states},
  author = {Haye Hinrichsen},
  journal = {Advances in Physics},
  volume = {49},
  issue = {7},
  pages = {815},
  numpages = {143},
  year = {2000},
  doi = {10.1080/00018730050198152}
}

@article{driven1,
  title = {Driven diffusive systems. An introduction and recent developments},
  author = {B Schmittmann, R.K.P Zia},
  journal = {Physics Reports},
  volume = {301},
  issue = {1-3},
  pages = {45},
  numpages = {19},
  year = {1998},
  doi = {10.1016/S0370-1573(98)00005-2}
}

@article{active1,
  title = {Novel Type of Phase Transition in a System of Self-Driven Particles},
  author = {Vicsek, Tam\'as and Czir\'ok, Andr\'as and Ben-Jacob, Eshel and Cohen, Inon and Shochet, Ofer},
  journal = {Phys. Rev. Lett.},
  volume = {75},
  issue = {6},
  pages = {1226--1229},
  numpages = {0},
  year = {1995},
  month = {Aug},
  publisher = {American Physical Society},
  doi = {10.1103/PhysRevLett.75.1226},
  url = {https://link.aps.org/doi/10.1103/PhysRevLett.75.1226}
}

@article{active2,
  title = {Hydrodynamics of soft active matter},
  author = {Marchetti, M. C. and Joanny, J. F. and Ramaswamy, S. and Liverpool, T. B. and Prost, J. and Rao, Madan and Simha, R. Aditi},
  journal = {Rev. Mod. Phys.},
  volume = {85},
  issue = {3},
  pages = {1143--1189},
  numpages = {0},
  year = {2013},
  month = {Jul},
  publisher = {American Physical Society},
  doi = {10.1103/RevModPhys.85.1143},
  url = {https://link.aps.org/doi/10.1103/RevModPhys.85.1143}
}

@article{surface1,
  title = {Dynamic Scaling of Growing Interfaces},
  author = {Kardar, Mehran and Parisi, Giorgio and Zhang, Yi-Cheng},
  journal = {Phys. Rev. Lett.},
  volume = {56},
  issue = {9},
  pages = {889--892},
  numpages = {0},
  year = {1986},
  month = {Mar},
  publisher = {American Physical Society},
  doi = {10.1103/PhysRevLett.56.889},
  url = {https://link.aps.org/doi/10.1103/PhysRevLett.56.889}
}

@article{ts21,
  title = {Nonequilibrium phase diagram of Ising model with competing dynamics},
  author = {Gonzalez-Miranda, J. M. and Garido, P. L. and Marro, J. and Lebowitz, Joel L.},
  journal = {Phys. Rev. Lett.},
  volume = {59},
  issue = {17},
  pages = {1934--1937},
  numpages = {0},
  year = {1987},
  month = {Oct},
  publisher = {American Physical Society},
  doi = {10.1103/PhysRevLett.59.1934},
  url = {https://link.aps.org/doi/10.1103/PhysRevLett.59.1934}
}

@article{ts22,
    author = {Rieder, Z. and Lebowitz, J. L. and Lieb, E.},
    title = {Properties of a Harmonic Crystal in a Stationary Nonequilibrium State},
    journal = {Journal of Mathematical Physics},
    volume = {8},
    number = {5},
    pages = {1073-1078},
    year = {1967},
    month = {05},
    issn = {0022-2488},
    doi = {10.1063/1.1705319},
    url = {https://doi.org/10.1063/1.1705319}
}

@article{mult1,
    author = {Michael M. Danziger and Ivan Bonamassa and Stefano Boccaletti and Shlomo Havlin},
    title = {Dynamic interdependence and competition in multilayer networks},
    journal = {Nature Physics},
    volume = {15},
    number = {2},
    pages = {178-185},
    year = {2019},
    doi = {10.1038/s41567-018-0343-1},
    url = {https://doi.org/10.1038/s41567-018-0343-1}
}

@article{mult2,
  title = {Collective Phenomena Emerging from the Interactions between Dynamical Processes in Multiplex Networks},
  author = {Nicosia, Vincenzo and Skardal, Per Sebastian and Arenas, Alex and Latora, Vito},
  journal = {Phys. Rev. Lett.},
  volume = {118},
  issue = {13},
  pages = {138302},
  numpages = {6},
  year = {2017},
  month = {Mar},
  publisher = {American Physical Society},
  doi = {10.1103/PhysRevLett.118.138302},
  url = {https://link.aps.org/doi/10.1103/PhysRevLett.118.138302}
}

\newpage
\clearpage
\onecolumngrid

\section*{Supplemental Material}

\maketitle

To corroborate the results presented in the main text, we provide here additional results that further support some of the conclusions discussed there. In particular, we focus on the nature of the phase transitions observed in the low-temperature regime.

In the main text, the continuous behavior of the order parameter curves shown in Fig.~3 already suggests the presence of second-order phase transitions. Additionally, no signatures of phase coexistence were observed, such as hysteresis or multiple peaks in the order-parameter probability distribution.

\begin{figure}[H]
\begin{centering}
\begin{centering}
\includegraphics[scale=0.52]{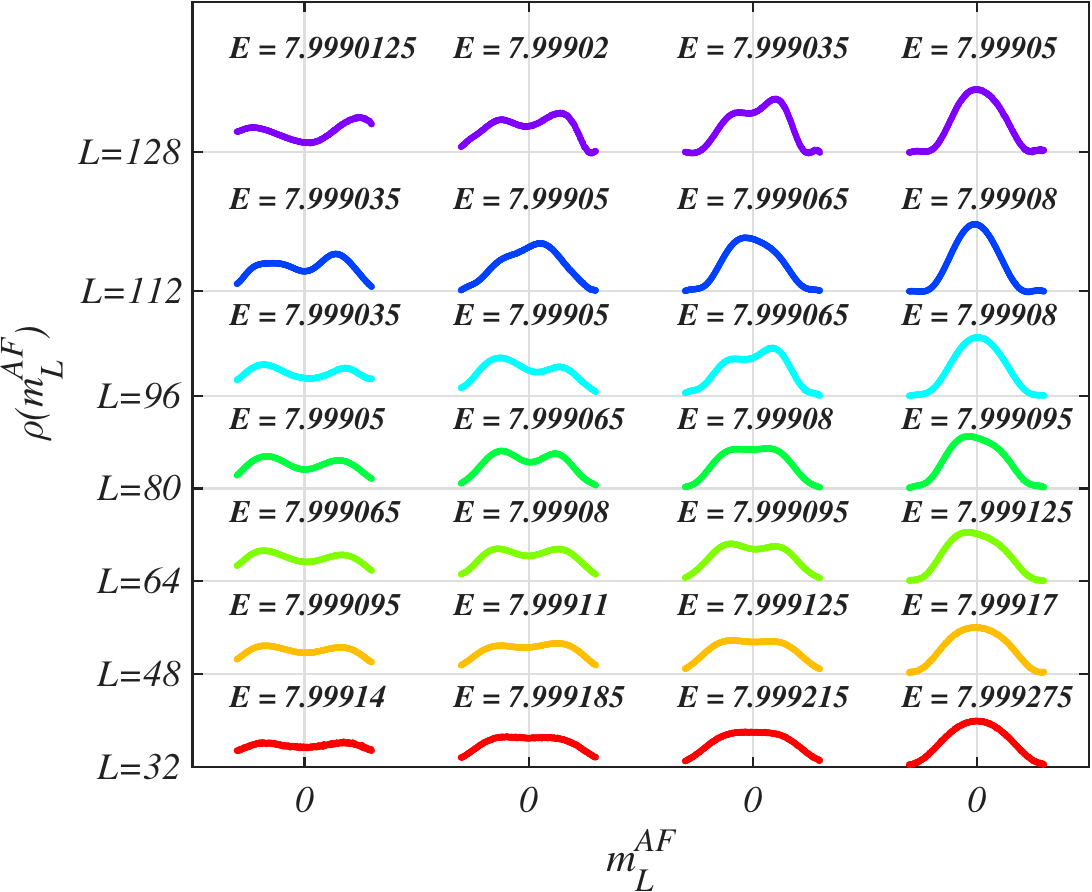}\hspace{0.25cm}\includegraphics[scale=0.52]{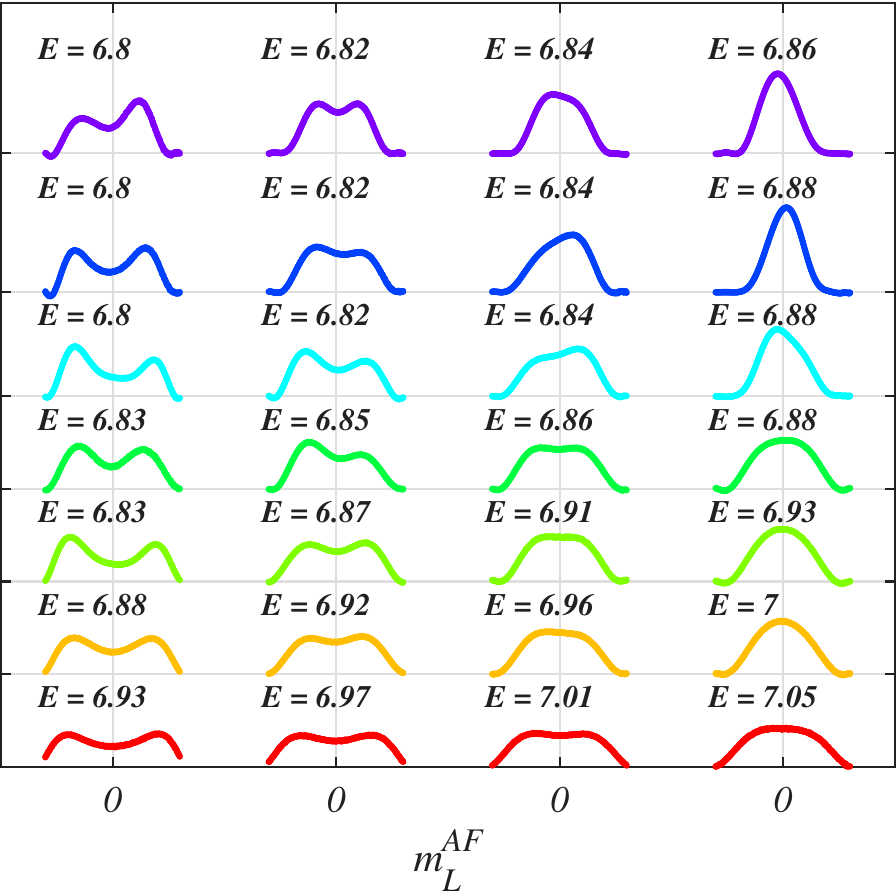}
\end{centering}
\caption{\textcolor{blue}{(Color online)} Probability distributions of the staggered order parameter $m_L^{AF}$ for different system sizes (indicated along the vertical axis) and different values of the drive $E$ (shown in the panels). The left column corresponds to $T=0.001$, while the right column shows the results for $T=1.0$.}
\label{fig:6}
\end{centering}
\end{figure}

To further confirm the absence of phase coexistence between the antiferromagnetically ordered phase and the paramagnetic phase, Fig.~\ref{fig:6} of this Supplemental Material shows the probability distributions of the staggered order parameter for two representative temperatures with distinct critical behavior, namely $T=0.001$ and $T=1.0$.

For low values of the drive $E$, and close to the critical point, the distributions exhibit two symmetric peaks at nonzero values of $m_L^{AF}$, reflecting the symmetry of the antiferromagnetically ordered phase. As the drive increases, these peaks progressively approach one another and eventually merge into a single peak centered at $m_L^{AF}=0$, indicating the emergence of the disordered phase above the critical point.

\begin{figure}
\begin{centering}
\begin{centering}
\includegraphics[scale=0.38]{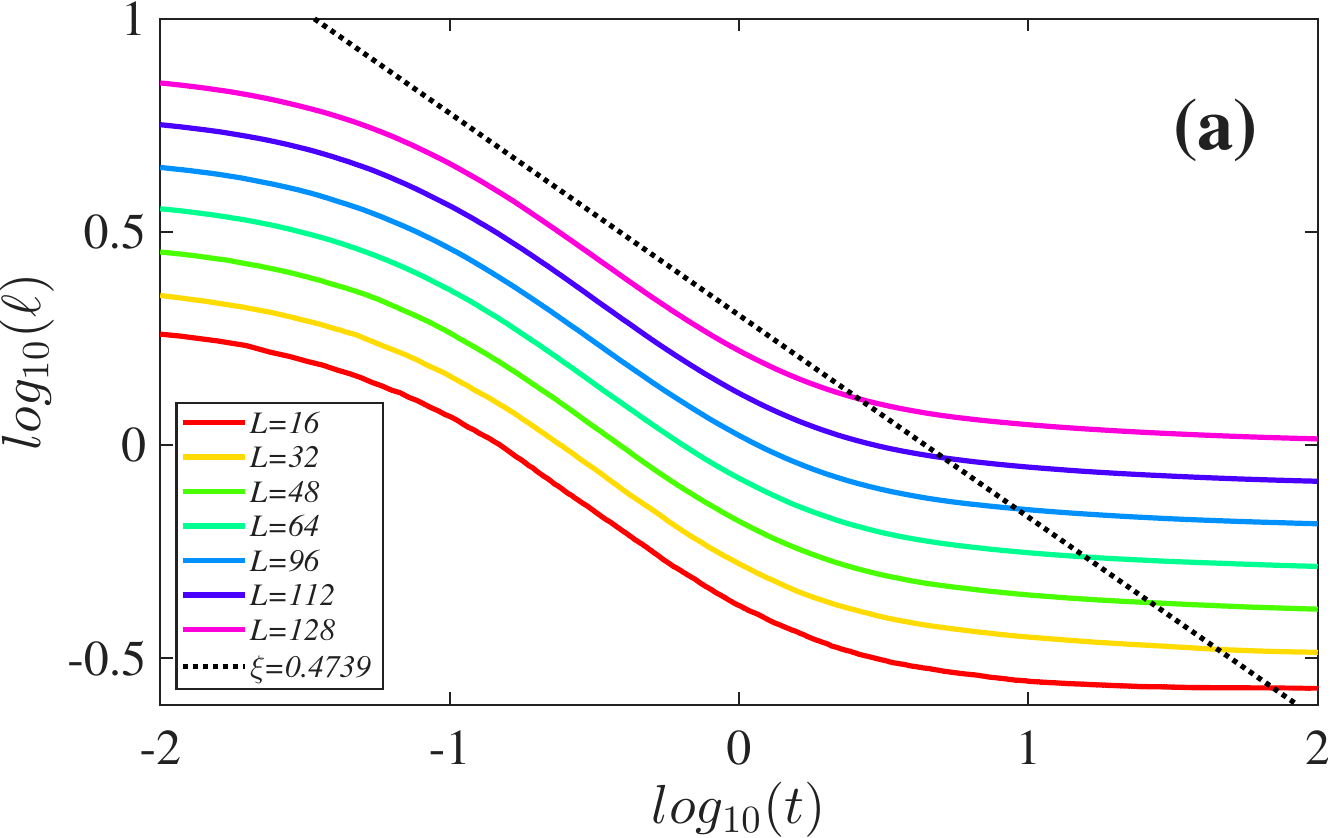}\includegraphics[scale=0.38]{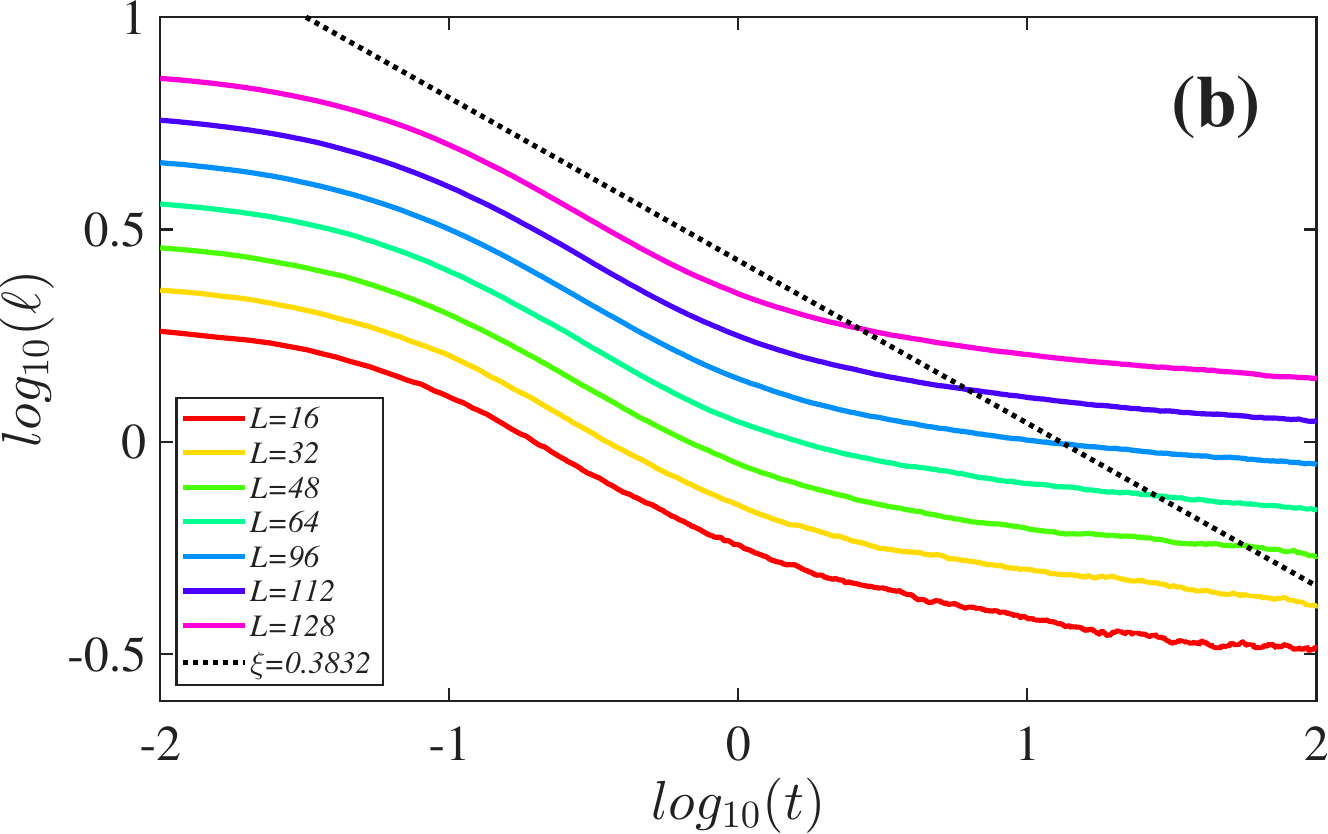}
\includegraphics[scale=0.38]{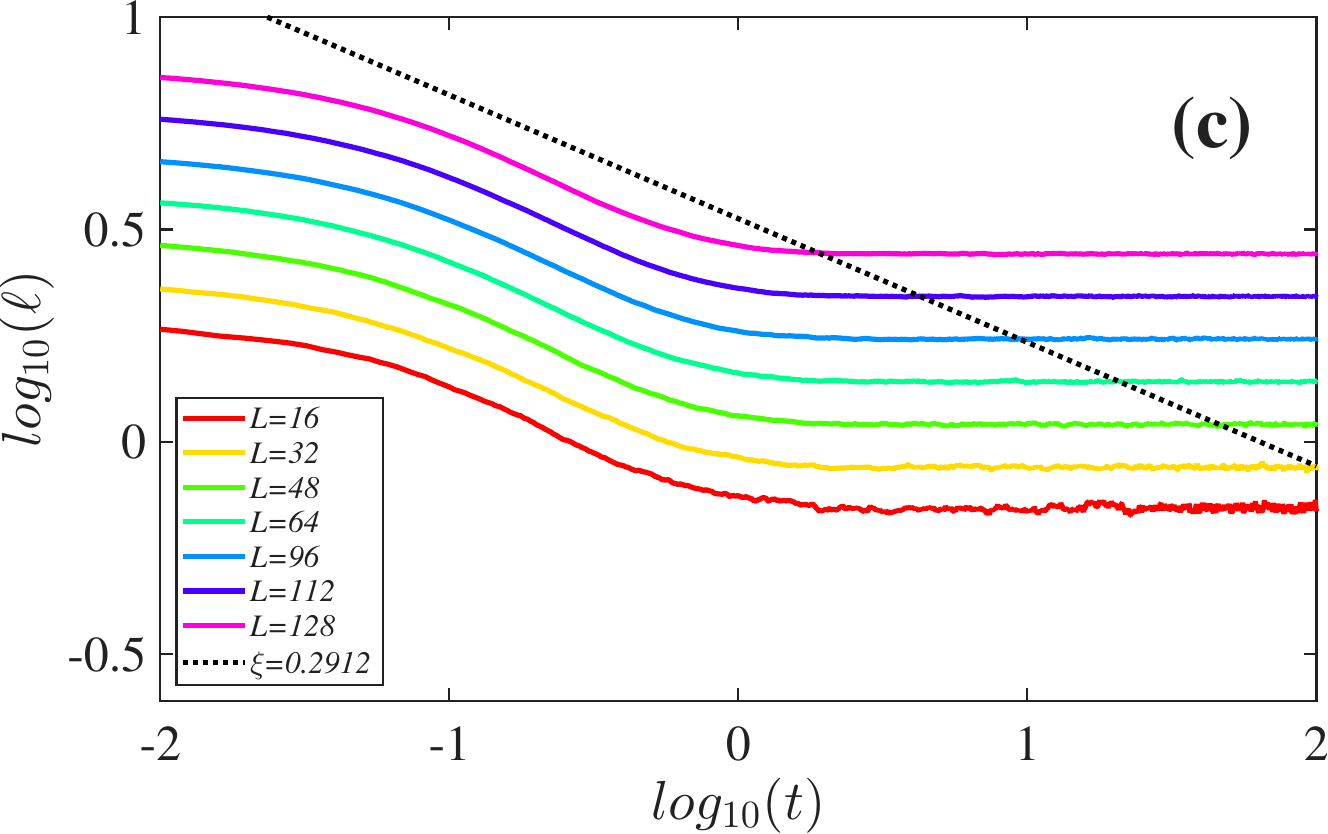}
\end{centering}
\caption{\textcolor{blue}{(Color online)} Log--log representation of the average domain size $\ell(t)$ versus time, employed to estimate the dynamic exponent governing domain growth, $\xi$, at $T = 0.01$. (a) Different network sizes for $E = 1.0$. (b) Field value near the critical point, $E = 7.9902$. (c) Behavior for $E = 10.0$. For clarity, the curves were vertically shifted by a factor of $0.1$ between successive system sizes, from the largest to the smallest, except for $L=128$.
}
\label{fig:7}
\end{centering}
\end{figure}

As an illustration of the behavior discussed in the main text regarding the exponent $\xi$, which characterizes the scaling of the average domain size $\ell(t)$ during the system evolution toward the stationary state, Fig.~\ref{fig:7} of this Supplemental Material shows the time dependence of $\ell(t)$. The curves display the evolution of domains until the stationary regime is reached, as well as the variation of the exponent $\xi$ across the three regimes considered here: before the critical field $E=1$, in the vicinity of the critical point $E_c$, and above the critical field $E=10$. 

In this example, we present the case $T=0.01$, but the same qualitative behavior is observed for the other temperatures discussed in the main text.

\end{document}